\documentstyle{l-aa}
\def\etal{{\it et al.\,}}
\def\eg{e.g.\,}
\def\ie{i.e.\,}
\def\lya{Ly$\alpha$\,}
\def\ze{$z_{em}$}
\def\za{$z_{abs}$}
\def\zae{$z_{abs} \approx z_{em}$}

\def\tone{\tau_{0.1}}
\def\tfive{\tau_{0.5}}
\def\ang{{\rm \AA}}
\def\mn{m_{\rm narrow}}
\def\BGE{\begin{equation}}
\def\EDE{\end{equation}}
\def\kms{km~s$^{-1}$}
\def\lyb{Ly$\beta$}
\def\nhi{$N$(HI)}
\def\cm2{cm$^{-2}$}
\def\ZnII{ZnII$\lambda\lambda$2025,2062}

\begin{document}
\input epsf
\thesaurus{03(11.01.1; 11.05.2; 11.09.3; 11.17.1; 11.17.4 Q0000--2619)}
\title{The metal systems in Q0000--2619 at high resolution
\thanks{Based on observations collected at the
European Southern Observatory, La Silla, Chile.}}
\author{Sandra Savaglio$^{1,2}$, Sandro D'Odorico$^1$,
\and Palle M\o ller$^{1,3}$}
\offprints{S. Savaglio}
\institute{$^1$ European Southern Observatory, Karl
Schwarzschild Stra\ss e, 2, W--8046 Garching bei M\"unchen, Germany \\
$^2$ Dipartimento di Fisica, Universit\`a della
Calabria, I--87036 Arcavacata di Rende, Cosenza, Italy \\
$^3$ Space Telescope Science Institute, 3700 San Martin Drive,
Baltimore, MD 21218. Affiliated with the Space Science Department, ESA}

\date{Received date; accepted date}
\maketitle
\begin{abstract}

We have obtained high, 11 and 14 \kms, and medium, 40 and 53 \kms,
resolution spectra of the $z_{em} = 4.11$ quasar Q0000--2619 covering
the range 4400 \AA\ to 9265 \AA . We identify nine metal
absorption systems, of which four were previously known. A fifth
previously suggested system at $z_{abs} \approx 3.409$ (Turnshek \etal~
1991) is ruled out by our data. Two of the eight systems
for which the \lya~ line is in the observable range have a damped
\lya~ line. Six of the nine systems show evidence for complex
sub--component structure. At our resolution and S/N we identify a
total of 21 sub--components in the nine systems.
Five of the nine systems (11 of the 21 components) fall within the
$\pm 5000$ \kms~ range of the emission redshift, and are hence
classified as \zae~ absorbers.

Data for all lines redward of \lya emission, plus lines in the Lyman
forest belonging to one of the nine metal systems are given. By
modelling the absorption lines with Voigt profiles, we have
derived
estimates for the column densities and the $b$ values of the
intervening clouds. Available photoionization models have then been
used to derive the degree of ionization and the metal abundances.

For the two damped systems we find metal abundances of $\leq 1$\% and
$\leq 8$\% of solar values at redshifts of 3.0541 and 3.3901
respectively. These upper limits are
consistent with what would be expected from previous determinations at
lower redshifts, and our data are hence compatible with earlier
conclusions that no evidence is yet found for chemical evolution of
intervening damped and Lyman limit absorbers.

For the \zae~ systems we found indications of metallicities comparable to, and
even in excess of solar values.  These much higher values compared to the
damped
systems, are in favour of the intrinsic hypothesis for these systems.

\end{abstract}

\section{Introduction}\label{S1}

The discovery of quasars with redshifts in excess of four offers a
possibility to study not only the nature and evolution of the quasar
population itself (Warren \etal~ 1987; Schneider
\etal~ 1991; Schmidt \etal~ 1991; Irwin \etal~ 1991; Warren \etal~
1991), but also, through analysis of the absorption
lines in the quasar spectra, to extend our knowledge of the intervening
matter out to the redshift of the quasar.

Pressing the redshift limit further and further back towards the early
universe, we may expect to reach a point beyond which the
stellar--induced chemical evolution and enrichment are negligible.
Determining when this happened
is of major importance for our understanding of the formation and
evolution of galaxies. In fact negative evolution of the number density
of CIV absorption line doublets in quasar spectra at redshifts larger
than 1.5 has been found by Sargent \etal~
(1988), and by Steidel (1990a) who interpreted this as evidence
for chemical evolution of the absorbers. This
result is still controversial however, since it was obtained from the
number counts of CIV doublets only, not from actual measurements of
the abundances of individual systems. Converting the number counts into
abundances is not a trivial task, since both the CIV lines and the \lya
lines often are strongly saturated.

Abundances can be obtained from determination of column densities
of several different ions of the same element, and a comparison of
the results with model calculations.
Bergeron \& Stasi\'nska (1986, hereafter BS) discussed the physical
properties of metal systems in QSO spectra using their photoionization
models, for a grid of metallicities, HI column densities, and ionization
parameters. They found that systems with
low ionization and with $N$(HI) $ > 10^{20}$ \cm2 had metal
abundances a few tenths of solar.
Another way of obtaining metallicities, is from the measurement of ions
for which the ionization potential is higher than that of HI (e.g. ZnII
and CrII). Such elements will in HI regions exist
in predominantly one ionization stage, and will hence present a
possibility of obtaining a metallicity determination which is
model--independent. Metal abundances in damped systems mainly based on ZnII
observations were obtained by Meyer \etal~(1989), Pettini \etal~(1990) and
Meyer \& Roth (1990); the systems they observed are distributed in the
redshift interval 2--2.8 and all show metal abundances significantly lower
than the solar value.
Rauch \etal~(1990) determined the metallicity of the damped system at
\za~$=2.0762$ in the \ze~= 2.55 quasar Q2206--199N. They found a low
metallicity compared with solar, [Si/H] $\leq -2.4$, and they found that
C/Si and O/Si ratios are consistent with being solar.
Observations of SiII, CrII, and AlIII belonging to the
damped system at \za~= 1.920 in the same quasar (Bergeron \& Petitjean,
1991) gave Si/H about 5\% of solar, and a low dust content.
Steidel (1990b) determined metallicities of 8 Lyman limit systems with
$2.90 \le $\za$ \le 3.23$ using a grid of models calculated from
Ferland's CLOUDY photoionization programme (Ferland, 1988), and
found metal abundances in the range $-3.0 \le$ [M/H] $\le -1.5$. Also
in this study it was found that relative abundances of C, O, and Si
were consistent with being solar.
M{\o}ller \etal~(1993) analysed four systems, one damped
and two \zae~ in Q2116--358, and another \zae~ in Q0347--241.
They found a metal abundance of $3 \pm 2$\%
of solar for the damped system, but abundances close to or even in
excess of solar for the \zae~ systems. They also found possible
evidence for non-solar relative abundances of C, and N in
\zae~ systems.

To obtain absorption spectra of sufficient quality to perform a detailed
abundance analysis, the underlying quasar naturally has to be fairly bright.
The $z=4.11$ quasar Q0000--2619 (Webb \etal~ 1988) offers a special opportunity
in this respect, being the brightest $z > 4$ quasar known.

Webb \etal~ obtained both high and low resolution spectra of
Q0000--2619 with the 3.6m ESO telescope and reported the identification
of a damped \lya~system at \za~$= 3.39$, and a $z_{abs} > z_{em}$ metal
system at \za~= 4.13. Sargent \etal~(1989) found Lyman limit absorption
at \za~= 3.412 but correcting this for the mean shift of the Lyman limit
of about 6 \AA\ rest wavelength determined by Tytler (1982), the
redshift is shifted to \za~$= 3.38$, corresponding well to the damped
system at \za~$= 3.39$. Schneider \etal~(1989) confirmed the presence
of the damped system at \za~$= 3.39$ and Steidel (1990a) found another
metal system at \za~= 3.5363 of which he detected a very weak CIV
doublet and a \lya~line of moderate strength.
Lanzetta \etal~(1991) added a second suspected damped system at \za~=
3.053, discovered from a reanalysis of the Sargent \etal~(1989) data.
This system was confirmed by Levshakov \etal~(1992), who presented data
with 1 \AA~ resolution of the region in the Lyman forest between
4050 \AA~ and 4950 \AA . From their fit to the \lyb~ line they
obtained \nhi~$= 4.3 \pm 0.7 \times 10^{19}$ \cm2.
Turnshek \etal~(1991) found an emission line object at a distance of
19'' from Q0000--2619, and suggested that it could be a galaxy at
\ze~= 3.409; they also suggested that previously unidentified absorption
features in the published spectra of Q0000--2619 could be due to
absorption associated with the object. Steidel \& Hamilton (1992)
obtained broad--band colours of the suggested high redshift galaxy,
and reported that a redshift of \za~= 0.438 was more likely. They also
found an object at a distance of 2.8'' from Q0000--2619,
which they suggested could be associated with the \za~$=3.39$ damped
absorber, but its redshift still awaits spectroscopic confirmation.

In this paper we present new high resolution observations on
Q0000--2619 and discuss the metal systems in the spectra. A catalogue
of lines in the Lyman$-\alpha$ forest and a discussion of the
\lya~lines will be given separately (D'Odorico \etal~ 1993). In
addition to the previously known metal absorption systems we report here the
identification of another five metal systems, and attempt to place
limits on the abundances of eight of the nine systems. The remaining
system is a low redshift MgII absorber, for which we have no \lya .
Most of the systems have many of the lines lying in the \lya~forest and
often reveal a complex structure at our high resolution. Even redwards
of the Ly$\alpha$ emission, absorption
lines from different systems are often blended.
Obtaining exact abundances is hence extremely complicated. Even so, our
high resolution in the Lyman forest, and our wavelength coverage from
4400 \AA\ to 9265 \AA , should make our absorption line lists, and our
fitted cloud parameters and systemic redshift list, valuable reference
points for future studies which might reach higher signal to noise of
selected regions.

The paper is organized as follows. In section \ref{S2} we describe the
observations and the data reduction, in section
\ref{S2b} the method of analysis, in section
\ref{S3} the data on metal systems and the results from our model fits
to the data are presented for the different systems in order of
increasing redshift. Our results from profile fits to
all metal systems are summarized in Table 11. Section \ref{S5}
contains the conclusions of the paper.

\section{Observation and data reduction}\label{S2}

The observations of Q0000--2619 presented here were obtained July and
October 1990, during the commissioning phase of the ESO Multi Mode
Instrument (EMMI). This instrument is mounted at the Nasmyth focus of
the 3.5m NTT on la Silla, Chile, and is described in D'Odorico (1990).
We analyse here five spectra of Q0000--2619 covering in total the
spectral range from 4400 \AA\ to 9265 \AA.

\begin{table*}\caption[t1]{Log of the observations}
{\label{t1}}
\begin{center}
\begin{tabular}{ccccccccc}
\hline\hline&&&&&&&\\[-5pt]
ident & No & date  & range & FWHM & grating & CD    & slit & exposure\\
      & of spectra   &       & (\AA) & (\AA)&         & grism & width& (s) \\
[2pt]\hline&&&&&&\\[-8pt]
1 &  2  & 24--27/10/90 & $8270-9265$ & 2.8 & 7  & -- & 1.2'' & 4500/6000\\
2 &  1  & 21/10/90     & $7650-8155$ & 1.4 & 6  & -- & 1.2'' & 4500 \\
3 &  1  & 05/07/90 & $4400-7615$     & 0.8 & 9  & \#3 & 1'' & 7000 \\
4 &  2  & 15/10/90 & $4700-6452$     & 0.2 & 10 & \#5 & 1.2'' & 5400/6600 \\
5 &  2  & 18/10/90    & $5830-8550$     & 0.3 & 10 & \#4 & 1.2'' & 4200/5100\\
[2pt]\hline\end{tabular}\end{center}
\end{table*}

The red channel of the instrument was used in echelle mode with gratings
\#9 and \#10 (giving resolutions of 7700 and 28000), and in long
slit mode with grating \#7 (resolution 2600) and grating \#6
(resolution 5200). For the echelle observations,
a standard grism of EMMI is mounted with 90$^o$ rotation with
respect to the standard orientation and act as cross disperser.
Table~\ref{t1} contains a full log of the observations.
The second column gives the number of individual spectra
obtained with each setting. In cases where two spectra had been obtained
with the same setup, they were averaged using a sigma rejection
algorithm which automatically located and rejected cosmics.

The detector used in the red arm was ESO CCD\#18, a Thomson TH31156 with
$1024\times 1024$ 19 $\mu$m square pixels. In the red spectral region
the NTT+EMMI combination used for these observations, is probably among
the most powerful available to observers of faint sources at high
spectral resolution. This is due to a measured overall efficiency
around 7\% at the top of the blaze of the echelle orders (including the
effects of the atmosphere at 1 airmass, the telescope, the instrument
and the detector), to the very good average seeing at the NTT which
reduces slit losses, and to the low read--out--noise of the detector
(5 $e^-$/pixel). The seeing during observations was typically between
0.8 and 1.2 arcsec. It was measured from the FWHMs of stellar images
in the QSO field.

The CCD used for these observations is very uniform. For the echelle
spectra the flat field exposures obtained with a continuum lamp were
used only to monitor the uniformity of the response and to identify
the position of 3 bad columns in the different spectral orders.

The data were reduced with the ECHELLE and LONG SLIT software packages
available in MIDAS, the ESO data reduction systems, running on a SUN
workstation. The long slit spectra were extracted using the optimal
extraction option, with the sky spectrum determined over a $100-150$
pixel region.
Tracing of the orders in the echelle spectra was done on flat field
images. The QSO spectra were extracted as a simple sum of the signal
perpendicularly to the direction of the dispersion. The sky spectrum was
determined  below and above the object spectrum, for the echelle grating
\#9 sky was determined from 28 pixels, for the echelle grating \#10
from 4 pixels. For the grating \#10 spectra, in order to reduce the
noise added by the sky subtraction, the sky continuum was spline
interpolated between the strong sky emission lines. The cores of
saturated absorption lines show no discernible offset from the zero
level in the final spectra, confirming the overall accuracy of the
sky subtraction.

For the wavelength calibration Th--Ar lamp spectra were used. The $rms$
measured from arc lines was typically $3-9$\% of the FWHM, the only
exception being the R28000 echelle spectrum \#4, where the wavelength
calibration had an $rms$ of 14\% of the FWHM. The FWHM of the
resolution, listed in the fifth column of Table~\ref{t1}, is the
average FWHM of the lines of the Thorium--Argon calibration spectra
over the spectral range of the observation. An independent check of
the wavelength calibration of the
echelle spectra was obtained by measuring the position of some known sky
emission lines in the range 5300--6400 \AA, the $rms$ values found were
identical to the ones obtained from the arc spectrum. The FWHM was
sampled by two to three wavelength bins over the entire wavelength
range. Flux calibration was obtained using the spectrophotometric
standards Feige 110 (Stone 1977) and LTT7987 (Stone \& Baldwin 1983;
Stone \& Baldwin 1984).

Variance spectra, used to determine the errors on the equivalent widths
of the absorption lines, were obtained by propagation of the photon
statistics of the object and sky spectra, and from the detector
read--out--noise.

\section{Absorption line measurements and modelling}\label{S2b}

Determination of the continuum in the QSO spectrum is a critical step in
the measurement of the absorption line equivalent widths. While the
continuum redwards of \lya~emission can be determined with good
accuracy, the high line density in the Lyman forest makes this task
very difficult bluewards of \lya~emission.

The method of defining the continuum to be where the local variance is
consistent with photon noise statistics (Carswell \etal~ 1982) is not
useful at this high redshift, since there are virtually no regions
unaffected by absorption lines. We hence followed the pragmatic
approach of Hunstead \etal~(1986), where the QSO continuum level was
determined by connecting in a smooth way the highest points of our
flux calibrated spectrum. The condition to avoid discontinuities in
the continuum, has as consequence that only few points can be used.

An independent check of the adopted continuum in the Lyman forest, can
be obtained from comparison of the "flux deficit" ($D_A$), as defined
by Oke \& Korycanski (1982), in the region  from 1050 and 1170 \AA\
rest wavelength, and the line blanketing in the Lyman forest ($a_o$),
with published values.
Webb \etal~(1992) found $D_A = 0.52 \pm 0.01$ by extrapolation of the
continuum redwards of \lya~emission. Since $D_A$ measured in this way
contains both the contribution from the line blanketing ($a_o$), and the
one caused by a diffuse Gunn--Peterson medium (if present), the value we
would find, which is only due to the lines, should be
$a_o \le D_A = 0.52$. Webb \etal~ also determined the contribution
only due to the line blanketing, for which they found $a_o$ = 0.45.
Even though $a_o$ in itself is a well defined parameter, the value
actually measured will depend strongly upon the resolution of the
spectrum. Since our spectrum has much
higher resolution than the one published by Webb \etal , 11 \kms~
versus 30 \kms, we have in total the condition $0.45 < a_o \le 0.52$. We
measure from our spectrum $a_o = 0.52$, and conclude that the overall
accuracy of the fitted continuum is $\le 10\%$ in the forest.

A dedicated quasar spectrum reduction software was used to find,
identify, measure, and model the absorption lines. The programme
calculates the centroid position, FWHM, equivalent width ($W_o$),
and corresponding error ($\sigma (W_o)$) of the observed absorption
features once the line boundaries are defined for each of them. For
well defined unblended lines the boundaries could be defined by
continuum crossings, but in most cases, lines in the \lya~forest and
multicomponent metal absorption features, the boundaries between
lines were defined as being the local maxima.

The automatic search procedure detected all absorption features with a
significance of 4$\sigma$ or more. Our line list redwards of Ly$\alpha$
emission is hence complete to 4$\sigma$, and to this limit only one line
remains unidentified; namely the feature at 6273.2 \AA~ which appears
at 4.0 sigma and 2.8 sigma in the spectrum of 11 \kms~resolution, and
40 \kms~ resolution respectively. This line could be a
CIV(1548) line associated with the damped system at \za~= 3.054.
Bluewards of Ly$\alpha$ emission an absorption line can be found at
almost any wavelength, and it is almost invariably blended. In this
part of the spectrum we therefore only list lines if the major part
of the blend could be due to the system in question
because of the wavelength coincidence and observed equivalent width, while
blends which are clearly dominated by Lyman forest lines are ignored.

Line modelling was done by convolution of calculated Voigt profiles
with the instrumental profile. The transition probabilities and
oscillator strengths that are required for the programme were taken from
Morton (1991). In the case of the complex profiles observed
for some of the metallic lines, a minimum number of components needed
to obtain a good fit was deduced from
the CIV profiles in the high resolution spectra.
The model profiles are computed from an initial estimate of the
column density and the $b$--value and then repeated until it is judged
that a satisfactory fit is obtained.
A fitting procedure  based on an automatic minimization of
the $\chi^2$ could not be applied to our data, which include
 different lines of the same ion,
 multiple components and  spectra of different resolution and
signal--to--noise ratio. While our global eye--fitting procedure
does not give  unique results, we have estimated
and give in Table 11 the maximum ranges of column densities (within the
acceptable intervals of $b$) which gives residuals comparable with the
noise spectrum.

\begin{figure*}
\picplace{7.5 cm}
\caption[1]{Absorption lines in a portion of spectrum \#2 of Table 1. The
normalized spectrum is shown as a thick line, the model fit as a light line
and the noise per resolution element as a dotted line. The position of the
measured lines is marked with the ion identifications and the system
identification.}
\end{figure*}

\begin{figure*}
\picplace{7.5 cm}
\caption[2]{As in Fig.~1, but portion of spectrum \#3 of Table 1.}
\end{figure*}

\begin{figure*}
\picplace{22 cm}
\caption[3]{As in Fig.~1, but portion of spectrum \#4 of Table 1.}
\end{figure*}

\setcounter{figure}{2}
\begin{figure*}
\picplace{22 cm}
\caption[3]{--{\it Continued}.}
\end{figure*}

\begin{figure*}
\picplace{22 cm}
\caption[4]{As in Fig.~1, but portion of spectrum \#5 of Table 1.}
\end{figure*}

\section{Observed properties of the metal systems}\label{S3}

With the 4865 \AA~interval covered by our echelle and long slit spectra,
we were able to analyse the four absorption systems known previously,
and in addition to identify five new systems, found through the detection
of CIV doublets in the region close to the emission redshift. In
tables~\ref{t2c} through \ref{t7} we list all of the suggested line
identifications for all the nine systems.

\subsection{Method of abundance determination}

In the study of BS, expected column density ratios of various
ions are plotted versus the column density of neutral Hydrogen for
several values of the ionization parameter $U$, defined as
the ratio of the density of ionizing photons to the gas density.
Given the column density for two different ions of the same element,
and the column density of HI, one can hence read off the ionization
parameter. In cases where only one ion of an element was detected, the
non--detection of other ions was used to constrain the ionization
parameter. In cases where such constrains were insufficient, we
assumed the relative Nitrogen to Carbon abundance to be solar, and
constrained the ionization parameter under this assumption. Using
the ionization parameter thus determined, or constrained, and the column
density of CIV, the $\log N$(CIV) versus $\log $\nhi~plot from BS can
be used to find the Carbon abundance of
the cloud, assuming that the ionization parameter is the same for all
elements and ions. In the case of a system with multiple components,
distinct values of the ionization parameter were computed for each
component. The assumption of constant $U$ for the different ions and
elements in a given component is not necessarily fulfilled if the
absorption occurs in physically separated regions.

In one case, for the damped \lya absorber at \za~$= 3.390$, it was
possible to apply the ZnII--method
described by Pettini \etal~(1990). As mentioned
in Section 1 this method contains no model assumptions, and hence less
uncertainties. For this reason we put a special effort in the observations
of the spectral region where the redshifted ZnII would be expected.

\subsection{Presentation of the absorption system data}\label{S3a}

To facilitate the presentation of the data, we have assigned to
each system a letter, A to I, in order of increasing redshift.
Six of the nine systems have a multicomponent substructure. We have
not employed any velocity window, or other strict rules for the
classification into systems versus sub--systems. We have simply
considered those components for which the lines were so blended that
they had to be fitted simultaneously, as one system.

The absorption line lists for each system are given in Tables
\ref{t2c} to \ref{t7}. The first column gives the resolution of the
spectrum where the line was detected, column 2 the observed wavelength
(heliocentric, and corrected to vacuum), columns 3 and 4
the measured equivalent width ($W_o$) and corresponding one
sigma ($\sigma (W_o)$). Line lists are complete to $4\sigma$. Lines
marked with an asterix in column 3, are lines which do not meet the
$4\sigma$ requirement, but for which the detection is supported by
detection in other spectra, or by good agreement with expected redshift,
and line strength. Columns 5 and 6 give the redshift and the
suggested line identification respectively. Footnotes in column 6 are
given for lines lying at the same position as another line belonging to
a different system, and for which the identification of the line (or the
blend) hence is not clear; such lines will appear in more than one
table. Listed equivalent widths of lines lying in the \lya~forest
($\lambda < 6214$ \AA ) should be regarded as upper limits, because of
the high probability of their being blended with forest
lines. Measurements of the absorption lines in all available spectra are
given, with high resolution data at the top, low resolution data at the
bottom, of each table. Hence each line often appears more than once.

Figures 1 -- 4 shows absorption lines detected in all the
systems. The observed spectrum, calibrated to wavelengths in air,
is plotted as a full line; model fit as a
light line; and the noise per resolution element as a dotted line.
The model spectra were calculated simultaneously for all the lines of
a given ion, and identical cloud parameters were used for all the
spectra. Only the binning, and the resolution profile with which the
models were convolved, differed. The positions of the measured lines
are marked with the ion identifications
and the system identifications. Note that only lines which have been
measured, and hence appear in the tables, are marked, while fainter
computed lines appear in the model spectrum without an identification.

In Table 11 we list the cloud parameters used for the line modelling.
When in addition to a preferred value, two additional values of $b$ are
given, they represent the range of
values still compatible with the noise. In the cases where a range is
given for a column density, it represents the maximum variation
compatible with the noise and within the given interval of $b$.
Where the column density is
given only as an upper limit, the line is either non--detected,
clearly blended, or simply lying in the Lyman forest. For the
non--detections	the upper limits are $\approx 3\sigma$ limits, for
blends and lines in the forest we give the best fit values, but regard
them as upper limits.

\subsubsection{The metal system at \za~= 1.434  -- A }\label{S3.1a}

This system was found by the identification of the MgII doublet. In the
high resolution spectrum the lines split up in two components as listed in
Table \ref{t2c}. The reality
of the system is verified by detection, in both our high and low
resolution spectra, of two weak FeII lines. No other metal lines are
expected within our observed range.

\begin{table}\caption[t2c]{The system at \za~= 1.434  -- A}
{\label{t2c}}
\begin{tabular}{cccccc}
\hline\hline&&&&&\\[-5pt]
 FWHM  & $\lambda_{obs}$& $W_o$ & $\sigma(W_o)$ & $z_{abs}$ &ID\\
(\kms) & (\AA)          & (\AA)     &  (\AA)      &           \\
\hline&&&&&\\[-8pt]
 11 & 6296.70 & 0.09$^*$ & 0.04 & 1.4343 & FeII(2586) \\
 11 & 6328.47 & 0.08$^*$ & 0.03 &  1.4339  & FeII(2600)\\
 11 & 6329.36 & 0.09 & 0.02 &  1.4342  & FeII(2600)\\
\hline&&&&&\\[-5pt]
 14 & 6329.29  &  0.12$^*$  & 0.06  &   1.4342 & FeII(2600)\\
 14 & 6805.94  &  0.48  &  0.07 &   1.4339 & MgII(2796)$^1$ \\
 14 & 6806.89  &  0.54  &  0.09 &   1.4342 & MgII(2796)$^2$ \\
 14 &  6823.46 &   0.24$^*$ &  0.08 &  1.4339  & MgII(2803) \\
 14 &  6824.50 &   0.46 &  0.11 &  1.4343  & MgII(2803) \\
\hline&&&&&\\[-5pt]
 40 &  6329.05 &   0.36$^*$ & 0.14  &  1.4341  & FeII(2600)\\
 40 &  6806.11 &  1.81  & 0.29  &  1.4339  & MgII(2796)$^{1,2}$ \\
 40 &  6823.61 &  0.96$^*$  & 0.26  &  1.4339  & MgII(2803) \\
[2pt]\hline&&&&&\\[-8pt] \end{tabular}

\scriptsize

$^1$ \za~= 3.3887 CIV(1550) \\
$^2$ \za~= 3.3893 CIV(1550) \\
\end{table}

The MgII$\lambda2796$ lies in the middle of the strong CIV absorption
at \za~$= 3.39$ making line fitting difficult. We obtained an acceptable
fit for a column density $N$(MgII) $ = 1\times 10^{13}$ \cm2 for both
components, with $b=7$ \kms~and $b=10$ \kms~for the lower and the high
redshift component respectively. The line at 6824 \AA\ was suggested to
be due to CIV(1548) by Turnshek \etal~(1991), but our data rule out
their identification.

\subsubsection{The damped \lya~system at \za~= 3.0541  -- B}\label{S3.1}

Proposed by Lanzetta \etal~(1991) as a damped system, it shows strong
\lya~absorption at $\lambda \simeq 4928$ \AA~(\za~$\simeq 3.054$). Our
search for CIV absorption at this redshift, resulted in a
possible detection of a very weak doublet at \za~= 3.0541 (Table
\ref{t3}). The CIV doublet is fitted well by a column density
of $5\times10^{12}$ \cm2. All other expected metal lines are in the
forest. We find a good fit to a possible SiIV doublet at the
redshift of the CIV, and to two strong lines which could be
identified with  OI and CII.

\begin{table}\caption[t3]{The system at \za~= 3.0541  -- B}{\label{t3}}
\begin{tabular}{cccccc}
\hline\hline&&&&&\\[-5pt]
 FWHM  & $\lambda_{obs}$& $W_o$ & $\sigma(W_o)$ & $z_{abs}$ &ID\\
(\kms) & (\AA)          & (\AA)     &  (\AA)      &           \\
\hline&&&&&\\[-8pt]
   11   & $\sim 4928$& $> 30$ &  ...   &$\sim 3.054$& \lya \\
   11   & 5279.65 & 1.79      & 0.17   & 3.0545   &  OI(1302) \\
   11   & 5410.74 & 1.43      & 0.17   & 3.0544    &  CII(1334) \\
   11   & 5650.72 & 0.57      & 0.09   & 3.0544    &  SiIV(1393)\\
   11   & 5687.20 & 0.15$^*$      & 0.08   & 3.0543    &  SiIV(1402)\\
   11   & 6276.49 & 0.06$^*$      & 0.02   & 3.0541    &  CIV(1548) \\
   11   & 6287.08 & 0.03$^*$      & 0.02   & 3.0542    &  CIV(1550) \\
[2pt]\hline\end{tabular}
\end{table}

We have constrained the column densities of SiII, SiIII, and
NV by non--detections of the lines SiII(1260), SiIII(1206),
and NV(1242) respectively. Fitting to the damped
\lya~profile we get a column density \nhi~$= 1.5 \pm 0.5 \times 10^{20}$
\cm2~(see Fig.~5), which is higher than the value $4.3 \pm 0.7
\times 10^{19}$ \cm2 as found by Levshakov \etal~(1992).

\begin{figure*}
\picplace{8 cm}
\caption[5]{Model fitting of the damped \lya~absorption at \za~$=3.0541$
in a flux (arbitrary scale) calibrated spectrum. The  Voigt
profile  for $N$(HI) $=1.5\times10^{20}$ \cm2 is shown as a continuum
line, while the dotted line refers to
a Voigt profile  for $N$(HI) $=4.3\times10^{19}$ \cm2~(Levshakov \etal~
1992).}
\end{figure*}

The low column density of CIV can be caused either by a low
metallicity, or by a very low ionization of the system. The last
possibility is consistent with all observations except for the low upper
limits of the column density of SiII and SiIII. From [CII/CIV]
we get $\log U > -3.2$. Assuming relative abundances to be
solar we get a consistent fit to C, Si, and N with $\log U = -2.8$ and
$Z \approx 0.001Z_{\odot}$.
For the lowest possible ionization, $\log U = -3.2$, we can have as
much as [C/H] $= -2$ provided Silicon is underabundant by an order of
magnitude compared to Carbon. Using the HI column density found by
Levshakov \etal , these limits get a factor of 3 higher.

\subsubsection{The damped \lya~system at \za~= 3.390   -- C}\label{S7}

This is the most distant damped system known.
 The \lya~of the system is the strongest
absorption line in the spectrum of Q0000--2619, and because of its
width, and of the blending with neighbouring lines, it is difficult
to establish its exact redshift. Adopting the redshift of 3.392 reported
by Webb \etal~(1988), we determine a HI column density of
$3.0\times10^{21}$ \cm2. The low ionization metal lines associated  with
the system are, however, found at a mean redshift of 3.390 (see
below and Table~\ref{t10a}), and at this
redshift the data are fitted better by a profile with \nhi~$=
2.5\times10^{21}$ \cm2 (Fig.~6). Also for this system we note that our column
density is somewhat higher than the value,
\nhi~$=(1.5 \pm 0.5)\times 10^{21}$ \cm2 determined from the measured
\lyb~ equivalent width, reported by Levshakov \etal~(1992). The
Ly$\beta$/OVI emission line of the QSO, lying in the blue wing of the
damped absorption, was modelled by shifting and scaling the \lya~
emission line, and adding it to the interpolated continuum.

\begin{figure*}
\picplace{8 cm}
\caption[6]{Model fitting of the damped \lya~absorption at \za~$=3.390$
in a flux (arbitrary scale) calibrated spectrum.}
\end{figure*}

\begin{table}\caption[t10a]{The damped Lyman--$\alpha$ at \za~= 3.390  -- C}
{\label{t10a}}
\begin{tabular}{cccccc}
\hline\hline&&&&&\\[-5pt]
 FWHM  & $\lambda_{obs}$ & $W_o$ & $\sigma(W)$ & $z_{abs}$&ID \\
 (\kms)& (\AA)                & (\AA)     & (\AA)       &      &\\
[2pt]\hline&&&&&\\[-8pt]
 11    & 5225.94     & 2.39 & 0.15 & 3.3900 & SiII(1190) \\
 11    & 5236.29     & 0.09$^*$ & 0.06 & 3.3881 & SiII(1193) \\
 11    & 5237.15     & 0.11$^*$ & 0.06 & 3.3888 & SiII(1193) \\
 11    & 5238.86     & 1.58 & 0.10 & 3.3903 & SiII(1193) \\
 11    & 5296.38     & 5.62 & 0.48 & 3.3899 & SiIII(1206)$^1$\\
 11    & $\sim 5337$ & ...  & ...  & $\sim 3.39$ & \lya \\
 11    & 5453.78     & 0.15$^*$ & 0.09 & 3.3883 & NV(1242) \\
 11    & 5535.00     & 1.04 & 0.07 & 3.3914 & SiII(1260)\\
 11    & 5716.77     & 1.67 & 0.10 & 3.3902 & OI(1302)\\
 11    & 5726.26     & 0.86 & 0.08 & 3.3901 & SiII(1304) \\
 11    & 5727.75     & 1.44 & 0.09 & 3.3912 & SiII(1304) \\
 11    & 5860.35     & 0.45 & 0.10 & 3.3913 & CII(1334) \\
 11    & 6115.96     & 0.61 & 0.05 & 3.3881 & SiIV(1393)$^2$ \\
 11    & 6117.17     & 0.66 & 0.05 & 3.3890 & SiIV(1393)\\
 11    & 6119.43     & 2.59 & 0.07 & 3.3906 & SiIV(1393)\\
 11    & 6155.50     & 0.60 & 0.02 & 3.3881 & SiIV(1402) \\
 11    & 6156.59     & 0.82 & 0.03 & 3.3889 & SiIV(1402) \\
 11    & 6159.07     & 2.12 & 0.05 & 3.3907 & SiIV(1402)\\
\hline&&&&&\\[-5pt]
 14    & 6702.29 & 1.30 & 0.12 & 3.3900 & SiII(1526) \\
 14    & 6703.77 & 0.20$^*$ & 0.07 & 3.3910 & SiII(1526) \\
 14    & 6793.77 & 1.15 & 0.11 & 3.3882 & CIV(1548) \\
 14    & 6795.07 & 0.59 & 0.09 & 3.3890 & CIV(1548) \\
 14    & 6796.63 & 1.02 & 0.10 & 3.3900 & CIV(1548) \\
 14    & 6797.94 & 1.40 & 0.10 & 3.3909 & CIV(1548) \\
 14    & 6799.07 & 0.27 & 0.06 & 3.3916 & CIV(1548)\\
 14    & 6804.07 & 0.66 & 0.09 & 3.3881 & CIV(1550)\\
 14    & 6805.94 & 0.48 & 0.07 & 3.3887 & CIV(1550) $^3$\\
 14    & 6806.89 & 0.54 & 0.09 & 3.3893 & CIV(1550) $^4$ \\
 14    & 6808.17 & 0.56 & 0.08 & 3.3902 & CIV(1550) \\
 14    & 6809.06 & 0.54 & 0.08 & 3.3908 & CIV(1550) \\
 14    & 6810.00 & 0.32 & 0.08 & 3.3914 & CIV(1550) \\
 14    & 7061.26 & 0.83 & 0.11 & 3.3901 & FeII(1608)\\
 14    & 7334.78 & 0.92 & 0.14 & 3.3900 & AlII(1670)\\
\hline&&&&&\\[-5pt]
 40    & 6702.47     & 1.69 & 0.22 & 3.3901 & SiII(1526) \\
 40    & 6794.29     & 1.59 & 0.23 & 3.3885 & CIV(1548) \\
 40    & 6797.72     & 2.85 & 0.29 & 3.3907 & CIV(1548) \\
 40    & 6806.11     & 1.81 & 0.29 & 3.3888 & CIV(1550)$^{3,4}$ \\
 40    & 6809.12     & 1.89 & 0.24 & 3.3908 & CIV(1550) \\
 40    & 7061.44     & 0.61$^*$ & 0.19 & 3.3902 & FeII(1608) \\
 40    & 7334.50     & 1.06$^*$ & 0.58 & 3.3899 & AlII(1670) \\
[2pt]\hline&&&&&\\[-8pt] \end{tabular}
\scriptsize
$^1$ \za~= 4.1325 OVI(1031) ~~~~~~$^3$ \za~= 1.4339 MgII(2796) \\
 $^2$ \za~= 4.0692 SiIII(1206) ~~~~~$^4$ \za~= 1.4342 MgII(2796) \\
\end{table}

Contrary to the previous system, the CIV absorption in this system
is strong. Our high resolution spectrum show the complex
multi--component structure, with a total velocity span of 232 \kms,
as seen also in other damped systems. It is difficult to make out
exactly how many sub--components are present. From the CIV profiles
we identify in the higher dispersion spectra (Fig.~4) a minimum of five
components which give a fit of both
lines of the doublet compatible with the noise and with the
lower resolution, higher S/N spectrum (Fig.~2). The SiIV
absorption falls in a less--than--average crowded region of the \lya~ forest:
assuming five components at the same velocities as CIV, we find
two very likely identifications and three upper limits.
Subsequently we obtained best fits to the SiII, SiIII, OI, and CII lines,
using the redshift components found from
SiIV and CIV. Except for SiII$\lambda1526$
all these low ionization lines are in the forest and we can set mainly firm
upper limits to the column densities.
Turnshek \etal~(1989) have suggested that two phases of different
excitation are usually detected in damped
\lya~systems. The higher excitation phase shows broader components spread
over a larger velocity range than the lower excitation, the components
of the latter having also a lower velocity dispersion. Our data have
a resolution which is between 5 and 10 times better than that of
Turnshek \etal~ (1989) and show that the strongest CIV
(the third of the system) coincide with the strongest SiII
absorption. Our metal abundances is based on the column densities for this
main component because we expect that it contains the bulk of the HI.
Assuming relative abundances to be
solar, we can obtain an acceptable fit to the damped component with
$\log U \ge -2.9$ and $Z \le 0.08 Z_{\odot}$.

\begin{figure*}
\picplace{7.5 cm}
\caption[7]{Long slit spectrum in the range of ZnII doublet at
\za~$=3.3901$. The expected position of the absorption lines is
marked. At a different {\it y--axis} scale, the
night--sky spectrum with the strong emission lines is also plotted. }
\end{figure*}

In general
abundance determinations are complicated by the unknown degree of
depletion onto interstellar grains of the most common metals seen, and
by the necessity to determine the ionization parameter of the absorbing
cloud. The degree of depletion will depend upon the dust to gas ratio of
the absorber. In the local interstellar medium (ISM) depletion can be as
high as 2--3 orders of magnitude
(Morton, 1975; Cardelli \etal~ 1991; de Boer \etal~ 1985).
Previous work on dust--to--gas ratio determination in damped
\lya~systems at $z \sim 2-3$  suggests that in general these systems
have lower dust--to--gas ratios than typical values of the local ISM.
Meyer and York (1987) found both Ni and Cr to be much less depleted in
two damped systems toward PKS0528--250 at \za~= 2.811 and 2.142 than
in the local ISM. This is consistent with the results of
Pettini \etal~(1990) who, from measurements of weak absorption lines
of ZnII and CrII in the \za~= 2.3091 damped system in
PHL957, found a dust--to--gas ratio approximately one order of
magnitude lower than locally, and with the most recent determination
by Meyer and York (1992), who found  a dust--to--gas ratio of
$\sim5$\% at a redshift of 0.692. An independent determination
was obtained by Fall \etal~ (1989), who, from the
statistical distribution of the spectral slopes of QSOs with
and without damped systems, also found that the damped absorbers must
have a dust--to--gas ratio one--tenth of the local value.

As pointed out by Pettini \etal~(1990), Zn is depleted much less
than most other elements, and moreover ZnII has an ionization
potential higher than that of HI. The column density of ZnII in a damped
absorber is hence
almost unaffected both by presence of dust and radiation field,
and abundances inferred are therefore model independent. Hence,
if ZnII lines can be found, they serve as solid indicators of
metal abundances.

The most recent value of solar abundance of Zn found by
Aller (1987) is (Zn/H)$_{\odot} = 3.8 \pm 0.7 \times 10^{-8}$ and the
degree of depletion in the ISM found by Van Steenberg and Shull (1988)
from a sample with
\nhi~$= 10^{20} - 10^{22}$ cm$^{-2}$ is [Zn/H]$_{\rm{ISM}} = -0.23$.
As pointed out above we would, at high redshifts, expect a depletion
an order of magnitude less than this, making it totally negligible. For
the absorber at \za~= 3.3901
we would hence expect a column density of $9.5 \times 10^{13}$ cm$^{-2}$
if abundances were solar.

To search for absorption from ZnII, we obtained long slit spectra
of the spectral range in which the redshifted ZnII(2025) and Zn(2062)
lines of the \za~= 3.3901 absorber are expected. At this redshift, both
lines are in spectral regions dominated by strong night--sky lines and
therefore obtaining a good spectrum of the night--sky is critical. We
used EMMI in the long--slit mode for this, and special care was taken
in the optimization of the extraction of both quasar and night--sky
spectra. In Figure 7 we show the night--sky spectrum, superimposed on
the spectrum of the quasar, with the expected position of both ZnII
lines marked. As is seen the ZnII(2062) line cannot be constrained
to any useful limit due to the sky lines, but the ZnII(2025) line is
expected in a small line free region between two strong lines, and hence
our detection limit can be determined from photon statistics. We find
an equivalent width 3$\sigma$ upper limit of $W_o=0.43$
\AA. In Figure 8 we
have plotted the curve of growth for the Zn(2025) line, for a set of
velocity dispersion parameters ($b$). The horizontal dotted line marks
our 3$\sigma$ rest equivalent width upper limit. For a $b$ value of 20
\kms~ we find $\log N$(ZnII) $\leq 12.8$, and hence a 3$\sigma$ upper
bound on the metallicity of $Z / Z_{\odot} \leq 0.06$. For $b=10$
\kms~ we have $\log N$(ZnII) $\leq 12.9$, then the upper bound is
$Z / Z_{\odot} \leq 0.08$, consistent with what already found.

\begin{figure}
\picplace{5.5 cm}
\caption[8]{Curve of growth of ZnII for different values of the velocity
dispersion parameter $b$ (in \kms). The dotted line represents the
equivalent width (in \AA) upper limit for the ZnII(2025) absorption line. }
\end{figure}

\subsubsection{The metal system at \za~= 3.536 -- D}\label{S3.3}

This system was first reported by Steidel (1990a). Associated with the
strong \lya~absorption (\za~= 3.535, $W_o=4.90$ \AA), he
detected a weak CIV doublet ($W_o = 0.71$ \AA~and $W_o = 0.61$
\AA~respectively at a resolution of 60 \kms) at \za~= 3.5363.

In our higher resolution spectrum, the \lya~line is resolved into
two components with \za~= 3.5358 and \za~= 3.5374 (see Table~\ref{t5}),
while the corresponding \lyb 's are not within the observable spectral range.
The SiIV doublet is also clearly resolved into two components with
\za~= 3.5357 and \za~= 3.5369. The lines of the CIV doublet are at low
resolution only detected at 3 and 2 $\sigma$, but in our high resolution
the $\lambda1548$ line appears to have double structure consistent
with the structure seen for SiIV and \lya .

\begin{table}\caption[t5]{The metal system at \za~= 3.536  -- D}{\label{t5}}
\begin{tabular}{cccccc}
\hline\hline&&&&&\\[-5pt]
FWHM  & $\lambda_{obs}$ & $W_o$ & $\sigma(W_o)$& $z_{abs}$ & ID \\
(\kms) &  (\AA)               & (\AA)     & (\AA)      &           & \\
[2pt]\hline&&&&&\\[-8pt]
 11    & 5414.39 & 0.39$^*$      & 0.12       & 3.5374 & SiII(1193)\\
 11    & 5471.95 & 0.42$^*$      & 0.11       & 3.5354 & SiIII(1206)\\
 11    & 5473.58 & 0.49      & 0.11       & 3.5367 & SiIII(1206)\\
 11    & 5513.99 & 1.98      & 0.12       & 3.5358 & \lya \\
 11    & 5515.98 & 2.01      & 0.14       & 3.5374 & \lya\\
 11    & 5638.69 & 0.26$^*$      & 0.09       & 3.5371 & NV(1242)\\
 11    & 5719.26 & 0.71      & 0.07       & 3.5376 & SiII(1260)\\
 11    & 6054.67 & 0.16$^*$      & 0.06       & 3.5369 &CII(1334)\\
 11    & 6321.70 & 0.19      & 0.03       & 3.5357 &SiIV(1393)\\
 11    & 6323.34 & 0.14      & 0.03       & 3.5369 &SiIV(1393)\\
 11    & 6362.57 & 0.08$^*$      & 0.03       & 3.5357 & SiIV(1402)\\
 11    & 6364.22 & 0.12$^*$      & 0.04       & 3.5369 & SiIV(1402)\\
\hline&&&&&\\[-5pt]
 14    & 6323.46 & 0.17$^*$      & 0.07       & 3.5370 &SiIV(1393)\\
 14    & 7022.41 & 0.30$^*$      & 0.10       & 3.5358 &CIV(1548)\\
 14    & 7024.26 & 0.52$^*$      & 0.16       & 3.5370 &CIV(1548)\\
 14    & 7034.08 & 0.18$^*$      & 0.10       & 3.5359 &CIV(1550)\\
 14    & 7035.12 & 0.47$^*$      & 0.14       & 3.5368 &CIV(1550)\\
\hline&&&&&\\[-5pt]
 40    & 6322.35 & 0.22$^*$      & 0.12       & 3.5362 &SiIV(1393)\\
 40    & 7023.58 & 0.66$^*$      & 0.20       & 3.5366 &CIV(1548)\\
 40    & 7034.87 & 0.41$^*$      & 0.21       & 3.5364 &CIV(1550)\\
[2pt]\hline\end{tabular}
\end{table}

SiII in both components is constrained to $\log N$(SiII) $\le
13.0$ from the non--detection of the SiII$\lambda1526$ line.
CII is constrained in the low redshift component by the non--detection
of the CII$\lambda1334$ line to $\log N$(CII) $\le 13.2$, the
high redshift component coincides with a line in the forest, which
is fitted well by a column density of
$N$(CII) $= 2.2\times 10^{13}$ \cm2. The low redshift component of
NV is constrained to $\log N$(NV) $\le 13.4$ from the NV$\lambda1242$
line, while the
high redshift component could contain as much as  $\log N$(NV) $= 13.9$.

For the low redshift component we can constrain the ionization parameter
to $\log U \ge -1.5$ from $N$(SiIII)/$N$(SiIV) which is consistent with
$N$(CII)/$N$(CIV). Applying this condition we find a Carbon abundance
of about 4\% of solar, [C/H] $= -1.4$, and that Silicon is over--abundant
by about 2.5 $dex$ compared to Carbon, [Si/C] = 2.5. For the higher
redshift component we get $\log U \geq -2.7$ from $N$(CII)/$N$(CIV),
and $\log U \approx -2.0$ from $N$(SiIII)/$N$(SiIV), and hence a
metallicity of [C/H] $= -0.8$ and [Si/C] $= 1.4$. The uncertainty
in the column densities quoted in Table 11 corresponds to a maximum range
$-3.0<$ [C/H] $<-0.6$ and $-3.0<$ [C/H] $<-0.3$ in the
abundances of the two components
respectively. These error ranges are mainly determined by the large
span of uncertainty on the HI column densities.

\subsubsection{The metal system at \za~= 4.061  -- E}\label{S3.4}

In the work of Steidel (1990a) the absorption line at $\lambda \sim 7840$
was identified as SiII$\lambda1526$ of the strong metal system
at \za~= 4.13. This identification can be excluded by our non--detection
of the SiII$\lambda1260$ line. In this region our spectrum has a higher S/N
ratio. We find an additional line at 7849 \AA~ and hence
identify the two lines as the CIV doublet
at a redshift of 4.0615 (Fig.~1 and Table~\ref{t6}).

The \lya~lines of the system are completely merged and hence appear
as only one line in Table \ref{t6}. From the \lyb~and Ly$\delta$
lines, it can be seen that the HI absorption is indeed separated into two
components at \za~= 4.0605 and
\za~= 4.0618. A complex structure of the CIV doublet is also
suggested by our high resolution spectra, but the low S/N makes it
difficult to determine
the exact number of sub--components. We have assumed two components
as observed in \lyb, but probably the number is higher and
hence it is very hard to estimate  the
uncertainty
of the $b$ value and the lower limit to the column densities.
The CIV(1550) lines of this
system are also blended with the CIV(1548) of system F, an effect often
referred to as line--locking (Scargle \etal~ 1970; Wampler, 1991).
For the low redshift component, the non-detection of the
CII$\lambda1334$ line and the NV doublet
gives a lower and upper limit on the ionization parameter of $-3.0 <
\log U < -0.9$. The first limit infers a solar Carbon abundance,
and the second 1\% of solar.
For the high redshift component, the non-detection of the
CII$\lambda1334$ line and the NV doublet
gives corresponding limits of $-2.7 < \log U < -1.7$, giving Carbon
abundances between solar and 10\% of solar.

The two components have velocities relative to the
quasar of 2800 and 2900 \kms respectively, and they hence, like all
the systems discussed below, fall within the
definition of \zae~ absorbers, which are commonly defined as absorbers
inside the $\pm 5000$ \kms~ range of the emission redshift (Foltz \etal~
1986).

\begin{table}\caption[t6]{\label{t6}}
\begin{tabular}{cccccc}
\multicolumn{6}{c}{The metal system at \za~= 4.061  -- E} \\
\hline\hline &&&&&\\[-5pt]
 FWHM   & $\lambda_{obs}$ & $W_o$ & $\sigma(W_o)$& $z_{abs}$ & ID\\
 (\kms) &  (\AA)               & (\AA)     & (\AA)      &         \\
[2pt]\hline&&&&&\\[-8pt]
 11     & 4746.42 & 1.28 & 0.30 & 4.0612 & Ly$\epsilon$ \\
 11     & 4806.39 & 1.34      & 0.17 & 4.0607   & Ly$\delta$    \\
 11     & 4807.79 & 0.75      & 0.16 & 4.0622   & Ly$\delta$    \\
 11     & 5003.89 & 0.33$^*$ & 0.09 & 4.0616 & OI(988) \\
 11     & 5190.69 & 1.33 & 0.08 & 4.0605   & \lyb         \\
 11     & 5192.00 & 1.16 & 0.07 & 4.0618   & \lyb         \\
 11     & 5223.41 & 0.81 & 0.12 & 4.0618 & OVI(1031)\\
 11     & 6152.71 & 3.85      & 0.05 & 4.0612   & \lya \\
\hline &&&&&\\[-5pt]
 14     & 7836.27 & 1.01      & 0.19 & 4.0615   & CIV(1548) \\
 14     & 7849.16 & 1.48      & 0.24 & 4.0614  & CIV(1550)$^1$ \\
\hline &&&&&\\[-5pt]
 40     & 7054.58 & 0.27$^*$ & 0.21 & 4.0616 & SiIV(1393) \\
\hline &&&&&\\[-5pt]
 53     & 7836.42 & 1.10      & 0.12 & 4.0616   & CIV(1548) \\
 53     & 7848.99 & 1.18      & 0.14 & 4.0613  & CIV(1550)$^1$ \\
[2pt]\hline&&&&&\\[-8pt]\end{tabular}

\scriptsize

$^1$ \za~= 4.0697  CIV(1548)\\
\end{table}

\subsubsection{The metal system at \za~= 4.0688 -- F}\label{S3.4b}

This system shows weak CIV absorption associated with a strong
\lya~line (Table~\ref{t60}). The other metal lines are
in the \lya~forest, the SiIII detected is at the edge of the \za~=
3.3881 SiIV$\lambda1393$ absorption line.

\begin{table}\caption[t60]{The metal system at \za~= 4.0688 -- F}{\label{t60}}
\begin{tabular}{cccccc}
\hline\hline &&&&&\\[-5pt]
 FWHM   & $\lambda_{obs}$ & $W_o$ & $\sigma(W_o)$& $z_{abs}$ & ID\\
 (\kms) &  (\AA)               & (\AA)     & (\AA)      &         \\
[2pt]\hline&&&&&\\[-8pt]
  11 & 5199.24 &   1.65 &  0.12  &   4.0689 &  \lyb \\
  11 & 6115.96 &   0.61 &  0.05  &   4.0692 &  SiIII(1206)$^1$\\
  11 & 6161.98 &  2.23 &  0.04  &   4.0688 & \lya \\
\hline &&&&&\\[-5pt]
 14 &   7849.16  & 1.48 &  0.24 &   4.0699 &  CIV(1548)$^2$ \\
\hline &&&&&\\[-5pt]
 53 &   7848.99  & 1.18 &  0.14 &   4.0697 &  CIV(1548)$^2$ \\
 53 &   7860.18  & 0.50$^*$ &  0.14 &   4.0686 &  CIV(1550) \\
[2pt]\hline&&&&&\\[-8pt]\end{tabular}

\scriptsize

$^1$ \za~= 3.3881  SiIV(1393)\\
$^2$ \za~= 4.0613 CIV(1550)\\
\end{table}

As for the previous system, the
non--detection of the NV doublet and CII$\lambda1334$, compared
to the CIV column density, sets a range of the ionization
parameter $-2.7 < \log U < -1.7$, from which we deduce a
Carbon abundance of $-1.2<$ [C/H] $<0.0$.
The system has a velocity relative to the quasar of 2400 \kms, and
is possibly line--locked with system E.

\subsubsection{The metal system at \za~= 4.1010 -- G}\label{S3.5}

This system has a velocity of 500 \kms relative to the quasar. The
CIV detection is very weak (see Table~\ref{t6a}), but present in both
the higher and lower resolution spectrum. More uncertain is the
detection of the SiIV doublet.
The CII and SiIII lines are falling inside strong lines in the
forest and give only constrains to the lower limit of the ionization
parameter, while the upper limit is estimated from the non--detection
of the NV doublet. Also here the [Si/C] only gives a
consistent fit if Silicon is overabundant. For $-2.5<\log U < -1.5$, we get
$0.8 < $ [Si/C] $<2.7$ and $-0.6< $ [C/H] $ <0.3$.

\begin{table}\caption[t6a]{The metal system at \za~= 4.1010 -- G}{\label{t6a}}
\begin{tabular}{cccccc}
\hline\hline &&&&&\\[-5pt]
 FWHM   & $\lambda_{obs}$ & $W_o$ & $\sigma(W_o)$& $z_{abs}$ & ID\\
 (\kms) &  (\AA)               & (\AA)     & (\AA)      &         \\
[2pt]\hline&&&&&\\[-8pt]
 11 & 4845.02  & 0.80 & 0.14 & 4.1014 & Ly$\delta$ \\
 11 & 4961.26  & 0.80 & 0.10 & 4.1014 & Ly$\gamma$ \\
 11 & 5232.19  & 1.12 & 0.08 & 4.1010 & \lyb \\
 11 & 6200.97  & 2.08 & 0.03 & 4.1009 & \lya \\
\hline &&&&&\\[-5pt]
 14 & 7155.29 & 0.70 & 0.14 & 4.1008 & SiIV(1402)$^1$\\
 14 & 7897.57 & 0.50 & 0.11 & 4.1011 & CIV(1548)\\
 14 & 7910.38 & 0.37 & 0.09 & 4.1009 &  CIV(1550)\\
\hline &&&&&\\[-5pt]
 40 & 7110.16  & 0.38$^*$ & 0.22 & 4.1014 & SiIV(1393)\\
 40 & 7155.19  & 0.35$^*$ & 0.21 & 4.1008 & SiIV(1402)$^1$\\
\hline &&&&&\\[-5pt]
 53 & 7897.43 & 0.41 & 0.09 & 4.1010 & CIV(1548)\\
 53 & 7910.36 & 0.18$^*$ & 0.06 & 4.1009 & CIV(1550)\\
[2pt]\hline&&&&&\\[-8pt]\end{tabular}

\scriptsize

$^1$ \za~= 4.1338 SiIV(1393)\\
\end{table}

\subsubsection{The metal system at \za~= 4.126 -- H}\label{S3.6}

This system is proposed by the detection of CIV $\lambda1548$
as reported in Table \ref{t6c}. The CIV $\lambda1550$ line is
completely immersed in the $\lambda1548$ line of the strong
CIV doublet at \za~= 4.132, and it is hence another line--locking
candidate.
At the same redshift there is a very asymmetric \lya~profile similar
to that shown by the CIV line, which makes the system
identification quite realistic and suggests that it has more than one
component. The two probable components (relative velocities with
respect to the quasar of $-900$ \kms~ and $-1000$ \kms respectively)
both have very low HI column densities. Again the non--detection of
CII and NV constrains the ionization parameter to
$-2.7 \le \log U \le -1.7$ and $-2.9 \le \log U \le -1.2$ for the
\za~= 4.1260 and the \za~= 4.1270 system respectively.
The lack of useful constrains from other lines, makes it impossible
to get anything but limits on the Carbon abundances. We get for
the lower redshift system $0.2 \le$ [C/H] $\le 1.3$ and for the
higher redshift system $1.0 \le $ [C/H] $ \le 2.0$. This last result is
obtained
for a strongly blended, very low column density \lya~line, and must be
regarded as very uncertain.

\begin{table}\caption[t6c]{The metal system at \za~= 4.126 -- H}{\label{t6c}}
\begin{tabular}{cccccc}
\hline\hline &&&&&\\[-5pt]
 FWHM   & $\lambda_{obs}$ & $W_o$ & $\sigma(W_o)$& $z_{abs}$ & ID\\
 (\kms) &  (\AA)               & (\AA)     & (\AA)      &         \\
[2pt]\hline&&&&&\\[-8pt]
 11 & 6184.13 & 0.27 & 0.03  & 4.1257 & SiIII(1206) \\
 11 & 6231.58  & 1.49 & 0.05 & 4.1260 & \lya \\
\hline &&&&&\\[-5pt]
 14 & 7936.02 & 0.58 & 0.08 & 4.1260 & CIV(1548)\\
 14 & 7937.67 & 0.43 & 0.10 & 4.1270 & CIV(1548)\\
\hline &&&&&\\[-5pt]
 53 & 7935.86 & 0.71 & 0.12 & 4.1259 & CIV(1548)\\
[2pt]\hline\end{tabular}
\end{table}

\subsubsection{The metal system at \za~= 4.132 -- I}\label{S6}

This CIV system (Webb \etal~ 1988) is not only the strongest in
the Q0000--2619 spectrum, but also the most distant metal system known.
We identify (Fig.~4) five sub--components of this system with {\it infall}
velocities with respect to the quasar ranging from $-1200$ to
$-1400$ \kms. All five sub--components have extremely high ionization.

It is associated with a strong \lya~absorption centered at
$\lambda \sim 6237$ \AA, and with an equivalent width of
$W_o = 8$ \AA. In Table \ref{t7} we report the identification of the
other lines of the Lyman series down to Ly$\gamma$. In \lya~ all the
components merge into what appears as one single line, but from the
higher order Lyman lines we see the multiple structure of the system.
In particular the \lyb~line is clearly split in at least three
components. The HI column density is determined assuming that
there are five clouds as observed in CIV
and using simultaneous fitting to \lya~ and \lyb~(see Fig.~3).
The upper limits of the three central components are determined
by the observed \lyb~ profiles and due the
possible contamination of the \lyb~ by \lya~ at lower redshift.

\begin{table}\caption[t7]{The metal system at \za~= 4.132 -- I}{\label{t7}}
\begin{tabular}{cccccc}
\hline\hline &&&&&\\[-5pt]
 FWHM   & $\lambda_{obs}$ & $W_o$ & $\sigma(W_o)$& $z_{abs}$ & ID\\
 (\kms) &  (\AA)               & (\AA)     & (\AA)      &         \\
[2pt]\hline&&&&&\\[-8pt]
 11     & 4991.09 & 6.05      & 0.33 & 4.1320 &Ly$\gamma$ \\
 11     & 5262.02 & 2.32      & 0.17 & 4.1301 &Ly$\beta$ \\
 11     & 5264.41 & 2.05      & 0.16 & 4.1324 &Ly$\beta$ \\
 11     & 5266.57 & 1.92      & 0.16 & 4.1345 &Ly$\beta$ \\
 11     & 5296.38 & 5.62      & 0.48 & 4.1325 & OVI(1031)$^1$\\
 11     & 6192.22 & 0.16      & 0.02 & 4.1324 &SiIII(1206) \\
 11     & 6193.52 & 0.39      & 0.03 & 4.1335 &SiIII(1206) \\
 11     & 6194.60 & 0.28      & 0.02 & 4.1344 &SiIII(1206) \\
 11     & 6239.04 & 8.00      & 0.06 & 4.1322 &\lya \\
 11     & 6356.44 & 0.20      & 0.04 & 4.1310 &NV(1238) \\
 11     & 6358.27 & 0.19      & 0.03 & 4.1325 &NV(1238) \\
 11     & 6360.38 & 0.48      & 0.05 & 4.1342 &NV(1238) \\
 11     & 6380.84 & 0.12$^*$      & 0.06 & 4.1342 &NV(1242) \\
\hline &&\\[-5pt]
 14     & 6360.69 & 0.12$^*$ & 0.06 & 4.1345 & NV(1238) \\
 14     & 7150.24 & 0.29$^*$ & 0.12 & 4.1302 & SiIV(1393)\\
 14     & 7152.35 & 0.61 & 0.12 & 4.1317 & SiIV(1393)\\
 14     & 7155.29 & 0.70 & 0.14 & 4.1338 & SiIV(1393)$^2$\\
 14     & 7196.48 & 0.12$^*$ & 0.08 & 4.1302 & SiIV(1402)\\
 14     & 7199.17 & 0.20$^*$ & 0.12 & 4.1321 & SiIV(1402)\\
 14     & 7202.50 & 0.28$^*$ & 0.11 & 4.1345 & SiIV(1402)\\
 14     & 7941.78 & 0.41 & 0.08 & 4.1297 & CIV(1548) \\
 14     & 7943.80 & 1.26 & 0.11 & 4.1310 & CIV(1548) \\
 14     & 7945.87 & 1.42 & 0.11 & 4.1323 & CIV(1548) \\
 14     & 7948.39 & 2.21 & 0.14 & 4.1339 & CIV(1548) \\
 14     & 7955.08 & 0.21$^*$ & 0.09 & 4.1297 & CIV(1550) \\
 14     & 7957.09 & 1.31 & 0.15 & 4.1310 & CIV(1550) \\
 14     & 7959.17 & 1.16 & 0.13 & 4.1324 & CIV(1550) \\
 14     & 7961.74 & 2.51 & 0.21 & 4.1340 & CIV(1550)\\
\hline &&\\[-5pt]
 40     & 6359.06 & 0.26$^*$      & 0.25 & 4.1332 &NV(1238) \\
 40     & 7152.19 & 0.29$^*$      & 0.21 & 4.1316 &SiIV(1393) \\
40      & 7155.19 & 0.35$^*$      & 0.21 & 4.1338 &SiIV(1393)$^2$ \\
40      & 7196.95 & 0.27$^*$      & 0.21 & 4.1305 &SiIV(1402) \\
40      & 7201.47 & 0.64$^*$      & 0.29 & 4.1338 &SiIV(1402) \\
\hline &&\\[-5pt]
 53     & 7946.44 & 5.48      & 0.19 & 4.1327 &CIV(1548) \\
 53     & 7959.67 & 4.31      & 0.18 & 4.1327 &CIV(1550) \\
[2pt]\hline&&&&&\\[-8pt]\end{tabular}

\scriptsize

$^1$ \za~= 3.3899 SiIII(1206) \\
$^2$ \za~= 4.1008 SiIV(1402) \\

\end{table}

The $b$ parameters deduced from the fit of five components to the
CIV absorption complex are between 12 and 23 \kms, the total velocity
span is 263 \kms, and the total column
density of the whole complex is $N$(CIV) $= 1.1\times10^{15}$ \cm2.

We have reported in Table~10 a tentative identification of the OVI(1031)
line. The detection would be of particular interest to check
the high ionization level of this system as discussed by
Lu and Savage (1993). The line however falls in the wing of the damped \lya
at \za~$=3.39$ and would require observations at higher resolution and S/N
for a positive identification.

For the lower redshift CIV component, at \za~= 4.1297, the non--detection of
CII and NV gives $-2.8 \le \log U \le -1.7$ and hence
$-0.7 \le $ [C/H] $ \le 0.6$.
In the system at \za~= 4.1310, $N$(CII)/$N$(CIV) constrains the
ionization parameter to $\log U > -2.3$, and hence
[Si/C] $> 0.2$ and $0.1\leq$ [C/H] $\leq 0.8$.
The ionization parameter of the component at \za~= 4.1323 is
constrained by the non--detection of CII and SiIV to $-2.2 <\log U < -1.3$,
giving $-1.1<$ [N/C] $< -0.3$ and $-0.1<$ [C/H] $<0.5$.
For the two highest redshift
components at \za~= 4.1334 and \za~= 4.1342 the ionization parameter
is constrained by $N$(SiIII)/$N$(SiIV) to $\log U \simeq -1.7$ and
$\log U \simeq -1.5$ respectively,
from which we deduce [N/C] $\simeq -0.4$, [Si/C] $\simeq
1.5$ and [C/H] $\simeq 0.4$ in the first case, while [N/C] $\simeq
-1.0$, [Si/C] $\simeq 1.4$ and [C/H] $\simeq 0.1$ in the second.
The  uncertainties on the CIV,
SiIII and SiIV column densities determine a maximum range
 $0.1 \leq$ [C/H] $\leq 2.0$ and $-0.4 \leq$ [C/H] $\leq 2.6$  in the
abundances
of the two components respectively.

\begin{table*}\caption[t11]{Parameters of the components in the metal systems}
{\label{t11}}
\epsfysize=25cm
\epsffile{tab11.ps}
\end{table*}

\section{Abundances in quasar absorption systems}\label{S5}

We have determined, or constrained, metal abundances of two damped
intervening systems (B,C), of two components of an intervening optically
thin system (D), all at redshifts in excess of 3, and of a total of 11
components in 5 \zae~ systems, all at redshifts close to 4.1.
For many of the systems we have estimated the metal abundances.
The uncertainties on the column densities quoted in Table~11 and the
assumption of the models suggests that these measurements have to
be taken with caution, but it seems unlikely that systematic errors
are present. In one case, the damped system at \za~$=3.390$, the model
dependent value was found in good agreement with the more
straightforward estimate based on the upper limit of the ZnII
absorption lines.

We find that the two damped systems, both have
low metal abundances, less than 1\% and 8\% of solar respectively, and
the two components of the intervening optically thin system
both have [C/H] $\approx -1$, but [Si/C] $\approx 2$.
For the components of three \zae~ systems (E,F,G), we estimate [C/H]
abundances between solar and 1\% of solar. For the two remaining
\zae~ systems (H,I), the carbon abundances range from [C/H] $=-0.7$
to [C/H] $=2.0$. The data on the high ionization system I are of particularly
good quality. They indicate a significant difference with respect to the
abundances in the damped and intervining systems.
For some of the \zae~ systems we also find
that [Si/C] $> 0$ and [N/C] $< 0$.

M{\o}ller \etal~
(1993) determined metal abundances of one damped intervening
system, and 3 \zae~ systems, all at redshifts $z \approx 2$.
Comparing their results to previously published determinations, they
concluded that there is a fundamental difference between the
intervening and the \zae~ systems, and suggested that the high metal
abundance \zae~ systems probably are related to BAL type systems which
also have high metal abundances (Turnshek 1988). In one of
their \zae~ systems they also found evidence for the non--solar
relative N/C abundances we have reported above.
The fact that the abundances of \zae~ systems show no evolution between
redshifts of 1.9 and 4.1 is an indication that these systems
are intimately connected to the quasar phenomenon itself, rather than to
its environment.

Finding in three components of the system I an under--abundance of Nitrogen
compared to Carbon, means that we could have introduced a possible bias for the
systems where we have used the assumption of relative solar abundances of N and
C to constrain the ionization parameter.  This bias would work in the following
way.  If [N/C] is negative, this allows the ionization parameter to be higher
before it comes in conflict with the non--detection of NV.  This could hence
raise the upper limit on $\log U$ by up to $0.5-1$, if it has been determined
from the non--detection of NV.  Since the original upper limit in all cases is
already quite high, it does not change any conclusions.  The column density of
CIV has its maximum at $\log U = -1.7$.  Adopting the higher upper limits for
$U$, will hence have no effect on the deduced limits on [C/H].  It will however
have some effect on the limits on [Si/C].  For a higher upper limits of $U$, we
predict lower observed column densities of all SiII, SiII, and SiIV.  Hence we
get much higher upper limits on [Si/C] for all the systems where Silicon has
been detected.  Since the ones quoted are already very high, this does not have
any impact on our conclusions.

The determined [Si/C] values, are surprisingly high, $1-3$. While our
determinations do possibly include quite large errors, due to the
strong blending of lines, and assumptions adopted, we would expect
these errors to be randomly distributed.
One place where there might
have been introduced systematic errors, is in the assumed ionizing
power law spectrum used by BS. Another possible source of the high
[Si/C] values
could be systems of mixed ionization. If a cloud has a high and a low
ionized part, we could have have both strong SiIII and CIV absorption
at the same time. In such cases the [C/H] have been under
estimated, and the true [C/H] is higher than the values given. We
could hence make the high [Si/C] disappear, but the [C/H] would in
that case get even higher. It is not clear if this also could make
the low [N/C] disappear.

\section{Summary}\label{S6a}

We have presented high and medium resolution spectra of the
$z_{em} = 4.11$ quasar Q0000--2619. We identified a total of nine metal
absorption systems, of which four were previously known. A fifth
previously suggested system at $z_{abs} \approx 3.409$
is ruled out by our data. Two of the eight systems
for which the \lya~ line is in the observable range have a damped
\lya~ line. Six of the nine systems show evidence for complex
sub--component structure. At our resolution and S/N we identify a
total of 21 sub--components in the nine systems.

We have attempted to obtain abundances for all systems, except for
the lowest redshift system where \lya~ is not observable. These are
the highest redshift systems for which such
an analysis has so far been made. For the two damped systems we were only able
to obtain upper limits to the metal abundances; $\leq 1$\% and
$\leq 8$\% of solar values at redshifts of 3.0541 and 3.3901
respectively. These upper limits are
consistent with what would be expected from previous determinations at
lower redshifts, and our data are hence compatible with earlier
conclusions that no evidence is yet found for chemical evolution of
intervening damped and Lyman limit absorbers. Observations with higher
S/N ratios, of the ZnII lines of the $z_{abs} = 3.39$ system,
confirm our metallicity determination.
No evidence was found for any non--solar relative
abundances in those systems.

For the \zae~ systems we found metallicities comparable to, and even in excess
of solar values.  These values are much higher than for the intervening and
damped systems, and when combined with the results of systems at lower
redshift, suggest that there is no substantial chemical evolution of the
absorbing gas intrinsic to the quasars between redshifts of 2 and 4.  Possible
evidence was found for non--solar relative abundances of Si, C, and N in the
\zae~ systems.

\paragraph{Acknowledgements.}

Credit has to be given to the technical staff of ESO for the excellent
quality of the data on Q0000--2619 obtained during the commissioning of
EMMI at the NTT. The unique contribution of H. Dekker and J. L. Lizon to the
instrument set--up is gratefully acknowledged. We thank P. Molaro for
the reduction of one spectrum and useful discussions and
M. Pettini for suggesting the ZnII observations of the damped \lya~system.
S. Savaglio acknowledges an ESO studentship to work at ESO Garching and travel
funds provided by the Italian Ministero della Ricerca Scientifica e
Tecnologica.

\end{document}